%% file: paper.tex
\documentclass{chi2012}
\usepackage{times}
\usepackage{graphicx}
\pdfoutput=1
\usepackage{graphicx}
\usepackage{color}
\usepackage{url}
\usepackage[greek, english]{babel}
\newcommand{\comment}[1]{}
\definecolor{Orange}{rgb}{1,0.5,0}

\begin{document}

\setlength{\paperheight}{11in}
\setlength{\paperwidth}{8.5in}
\setlength{\pdfpageheight}{\paperheight}
\setlength{\pdfpagewidth}{\paperwidth}

 \toappear{Submitted for review}

\title{Your browsing behavior for a Big Mac: \\
      Economics of Personal Information Online  }
\numberofauthors{5}
\author{
  \alignauthor Juan Pablo Carrascal\\
    \affaddr{UPF and Telefonica Research}\\
    \email{jpcr@tid.es}
  \alignauthor Christopher Riederer\\
    \affaddr{Columbia University}\\
    \email{cjr2149@columbia.edu}
  \alignauthor Vijay Erramilli\\
    \affaddr{Telefonica Research}\\
    \email{vijay@tid.es}
  \alignauthor Mauro Cherubini\\
    \affaddr{Telefonica Research}\\
    \email{mauro@tid.es}
  \alignauthor Rodrigo de Oliveira\\
    \affaddr{Telefonica Research}\\
    \email{oliveira@tid.es}
}

\maketitle

\begin{abstract}
\input{abstract}

\end{abstract}

\keywords{Auction, browser plugin, economics, Experience Sampling, Personal information, Privacy}


\section{Introduction}
\label{sec:intro}
\input{intro}

\section{Related Work}
\label{sec:relwork}
\input{relwork}

\section{Methodology}
\label{sec:method}
\input{method}

\section{Auction and Survey results}
\label{sec:results1}
\input{results}

\section{Discussion}
\label{sec:discus}
\input{discuss}

\section{Implications for Design and Future Research}
\label{sec:design}
\input{implications}

\section{Concluding Remarks}
\label{sec:conc}
\input{conc}


\bibliographystyle{acm-sigchi}
\bibliography{privacy-all}
\balancecolumns
\end{document}

%% file: abstract.tex
Most online services (Google, Facebook etc.) operate by providing 
a service to users for free, and in return 
they collect and monetize personal information
(PI) of the users. This operational model
is inherently economic, as the ``good'' being 
traded and monetized is PI. This model 
is coming under increased scrutiny as online services are moving
to capture more PI of users, raising serious privacy concerns. 
However, little is known on how users valuate different types of PI while 
being online, as well as the perceptions of users with regards to
exploitation of their PI by online service providers.

In this paper, we study how users valuate different types of PI while being online, 
while capturing the context by relying on Experience Sampling. We were able to extract the 
monetary value that 168 participants put on different pieces of PI. We find that users 
value their PI related to their offline identities more (3 times) than their browsing behavior. 
Users also value information pertaining to financial transactions and social network interactions more 
than activities like search and shopping. We also found that while users are overwhelmingly in favor of 
exchanging their PI in return for improved online services, they are uncomfortable if these same providers monetize their PI.

%% file: intro.tex
A large part of the Internet economy 
operates by monetizing personal information (PI) of end-users, 
primarily via online advertisements. Online service providers like Google, Facebook etc. offer services for free, and in return, collect, aggregate and monetize PI. 
However, this monetization comes at the cost of erosion of privacy of end-users. 
Entities like Google etc. are aggressively collecting \textit{more} PI about the end-users, often
outside the scope of their application (Google via Doubleclick cookies, Facebook 
via their `Like' button etc.) and have been vocal about their 
dim view of online privacy~\cite{mark_fb, eric}.  At the same time, users are becoming more aware of
various privacy breaches~\cite{url:fbprivacy, url:what, url:scrapers, url:apple}, attracting the attention
of regulatory bodies as well~\cite{url:privacybill}. 

The ecosystem of service providers on one end and users on the other can be viewed as a two sided market~\cite{Parker2000},
where the `good' being traded is PI
of users. In such a system, it is easy for service providers to attach a value on each users' PI, based on the revenues
they can extract. However, for users to perform cost-benefit type analysis, where the cost
is loss of privacy, and the benefit is the service they obtain in return, it is important
that they first know the value of \emph{their} PI they are trading away. 
In this paper, we focus on understanding this value that users attach to their own PI\footnote{We focus on monetary value 
assigned by the user to their information, although one can imagine other notions of value and utility like satisfaction, happiness etc. We consider money as we are interested in the overall ecosystem of online services that hinges on monetizing PI. Secondly, 
money is a tangible concept and easier to express as opposed to user happiness. We will consider other notions of value in future work.}, specifically
while web-browsing that has been shown to have serious 
privacy implications~\cite{bala-wisp, bala-www}.

It is challenging to extract the value that users' put on their own PI. First of all,
the valuation could change based on \emph{context}. For instance, the value
that a user puts on the fact that she is searching for a restaurant can be different
than when she is searching for cancer drugs. Indeed, it can even change
between the type of interactions; social interactions can have a different valuation
from a financial transaction conducted online. Second of all, valuations
may depend on personal demographics; one's education levels, socio-economic status, age and gender. 
Past work done in this domain
has included valuating personal information like weight, age etc. ~\cite{Huberman2005} as well
as location information~\cite{Ross-location}, however they all rely on surveys and fail
to capture the context.

The main research question we deal with is ``what value do users associate with their PI'', more specifically
their web browsing behavior. In order to capture context we rely on Experience Sampling (rESM)~\cite{Cherubini:2009fu} and 
develop and deploy a browser plugin (Sec: Methodology) to `sample' what users are experiencing and obtain their
responses in context. We sample users on the different types of content/services they access (Social, Search, Finance etc.). 
We recruited 168 participants spanning a diverse range of demographics
and used a reverse second price auction to obtain an honest valuation for different types of PI.
Our main findings include that users value PI related to their offline identity -- age, gender, address and
economic status at \textgreek{\euro} 25 (median) and this value does not change
across different services. This value is higher than what users associate with their browsing history,
which is \textgreek{\euro} 7 (median). In terms of valuating service specific 
PI (photos uploaded to your social network, search keywords, online purchases etc.), users had
different valuations, with interactions on Social and Finance web-sites getting high valuations (\textgreek{\euro} 12, 15.5).
Interestingly, we see no difference between the valuations users put on one piece of PI as opposed to multiple pieces of the
same PI.

The second research question we address is to understand users' perceptions on the economic usage of their PI 
by online service providers. Once again, we use the same
methodology, and record the responses in context. Our main result is that while most users
have knowledge about their PI being monetized, and while they are comfortable with this PI being used to improve services,
they are overwhelmingly negative about their PI being monetized. This contrast can have important implications 
for design of new services, as well as for future research.  


%% file: relwork.tex
There has been a considerable amount of work done on how users valuate personal information
and privacy while considering psychological, social, economic and technical factors. 
We review work related to our research question 1 (RQ1) and the question 2 (RQ2).

\emph{RQ1: What monetary value do users attach on different types of PI while being online?}

Previous research has shown that valuation can depend on the type of information release, for instance
Huberman et al~\cite{Huberman2005} have reported that valuation of certain bits of PI like weight and age depends
on the desirability of those bits of information in a social context. Users attached low values to their 
PI if their respective values were between typical values or if the users came out as `positive'; (e.g low weight)
in a social context.  Likewise, valuation of location information has been found to depend
on factors like the distance traveled by the user and to a lesser extent who the users communicate with~\cite{Ross-location}.
The authors of~\cite{Ross-location} used a reverse auction mechanism to estimate minimum monetary value 
that participants (undergraduate students in Cambridge, UK) would accept to disclose constant location information 
towards a scientific experiment or for commercial use.  They report a median of 10 pounds, with a highly skewed distribution. 
Interestingly, the possibility of commercial use of the data increased the median by 10 pounds. 
A similar, and larger study (spread over 5 European countries) reaffirmed the median value of 10 pounds, and also
established that users \emph{factor} in diminishing returns of more information, and hence started asking for less~\cite{cvrcek2006}. 

In a survey that was part of a larger study~\cite{Gideon2006}, users expressed different concerns for different types of information -- 
sharing financial information as well as purchasing activity of goods like condoms were of high concern, while general interests
were rated low. Some demographic factors appear to influence valuation as well, for instance
there seems to be some correlation between privacy attitudes (hence valuation) and income levels; people
with low salaries seem less concerned about privacy~\cite{Acquisti2005}. 
Our work differs in multiple regards -- we focus on web browsing information of users that is of economic interest
to entities like Google etc. and such information raises privacy concerns~\cite{bala-www, bala-soups}. 
A related work looked into Americans' attitudes towards behavioral advertising~\cite{advert} -- which is one 
primary method of monetizing PI, and relied on surveys. Second, we 
study the effects of demographic information like age, gender, education levels and socio-economic factors on valuation
of one's PI. And lastly, while the above papers used extensive surveys to figure out different valuations, 
we use a methodology based on experience sampling to capture the context and obtain valuations in-situ. 

Another body of work that is related to monetary valuation of PI has to do with
studying the dichotomy that exists between willingness to pay (WTP) to buy privacy protection
and willingness to accept (WTA) to reveal PI. A difference between WTA and WTP can be 
indicative of an \emph{endowment} effect~\cite{Thaler1980}: people can place a higher value on an object that they own, in this case PI. 
The authors of~\cite{Grossklags2007} report that while people are generally willing to accept small 
amounts for some types of personal data (weight), there is wide gap between WTP/WPA for 
private data (e.g. revealing number of sexual partners). An updated study~\cite{Acquisti2009} 
used traceable gift cards given to users to reveal that people who started with 
a position of greater privacy protection were also likely to forego money to reveal PI. 
The difference between WTA/WTP seems to suggest that the way privacy choices are framed 
may affect decisions people make with regards to their PI. 
This topic was dealt with in~\cite{Braunstein2011}, where they asked the same set of questions to 
three groups of participants. The privacy awareness in the language used for the different groups 
was progressively increased. They found a relationship between users' answers and the wording of 
privacy-related questions. As the use of privacy-related language increases, participants tend to give 
more importance to private content, along with a decrease in the willingness to share personal content (e.g. purchase history). 
In our paper we do not deal with WTP vs WTA explicitly, instead we focus on extracting WTA for 
web-browsing, while capturing as much context as possible. We also consider the results of~\cite{Braunstein2011}, 
and design our experiments with neutral language, so as not to bias the user one-way or another.

\emph{RQ2: What are the perceptions of users vis-a-vis their PI being monetized, improving existing 
services and for personalized advertisements?}

A majority of the work done on understanding the awareness levels of users in terms of how their PI is exploited
and related privacy concerns has focused on how the actual behavior of people deviates from what they state. 
This deviation has been noted by~\cite{Jensen2005a} who also found that there is a difference between reported knowledge
and reality; in general people do not seem to know as much about privacy protection measures as they state. They also
report that surveys as a method should not be taken as indicative of users' actual behavior. The authors of~\cite{Berendt2005}
divide society into privacy fundamentalists, marginally concerned, and classified the rest between those who are identity concerned (PI
about email, address etc.) and those who are profile concerned (PI about hobbies, interest etc.). Acquisti studies the reasons
that affect people's behavior vis-a-vis privacy and reports bounded rationality as well as the practice of 
hyperbolic discounting~\cite{Acquisti2004-sec}; assigning a higher value to actions involving immediate gratification 
than those actions leading to long-term protection. 
In this work, we focus on understanding people's knowledge and perception of how their PI is exploited from an economic view-point,
and use experience sampling to capture the behavior and context. 

Another form of gauging awareness levels is to understand if users read online privacy policies and if they understand them. 
Jensen et al~\cite{Jensen2004} conducted an analysis of 64 privacy policies of high traffic and health-care websites, focusing 
on the use of policies, their readability (using the Flesch Reading Ease Score), the equivalence between their 
legibility and education levels required for reading it and the way the websites handle changes to the policies.
It was found that policies were very hard to parse and understand, pointing to simpler methods to convey the same information.

%% file: method.tex
To answer our research questions we employed a refined version of
the \textit{Experience Sampling Method} (\textit{i.e.}, rESM).
Experience Sampling Method involves asking participants to report on
their experiences at specific points throughout the day. The method
was originally developed in the psychology
domain~\cite{Barrett:2001ve} and recently adapted successfully in
many studies of Human-Computer Interaction \cite{Intille:2003ys,
Consolvo:2003zr, Iachello2006, FischerESMMobileHCI2009}. As
Cherubini et al highlighted in a previous
paper~\cite{Cherubini:2009fu}, the main advantage of ESM is its
ability to preserve the ecological validity of the measurements,
defined by Hormuth~\cite{Hormuth:1986vn} as: ``the occurrence and
distribution of stimulus variables in the natural or customary
habitat of an individual''. This method compares with recall-based
self-reporting techniques --although recall delay is kept minimal--
by ``beeping'' the participant in close temporal proximity to when a
relevant event was produced.
One of the drawbacks of the method is that often participants are
sampled at random times or with little knowledge of their
whereabouts and therefore the beeping might be invasive for many
participants. This is why in recent years some researchers have
proposed to \textit{refine} the method by modeling the participants'
context \cite{Froehlich:2007qm, Cherubini:2009fu,
Vastenburg:2010ly}. Refined --or contextual-- experience sampling
methods attempt to go one step further by only signaling users at
appropriate times or in the right context.

As a means to perform rESM, we instrumented the web browser of
participants with a plugin that was able to log the website the
participant was browsing and classify the website according to 8
categories. 


\begin{figure}[htbp]
    \centering
        \includegraphics[width=1\columnwidth]{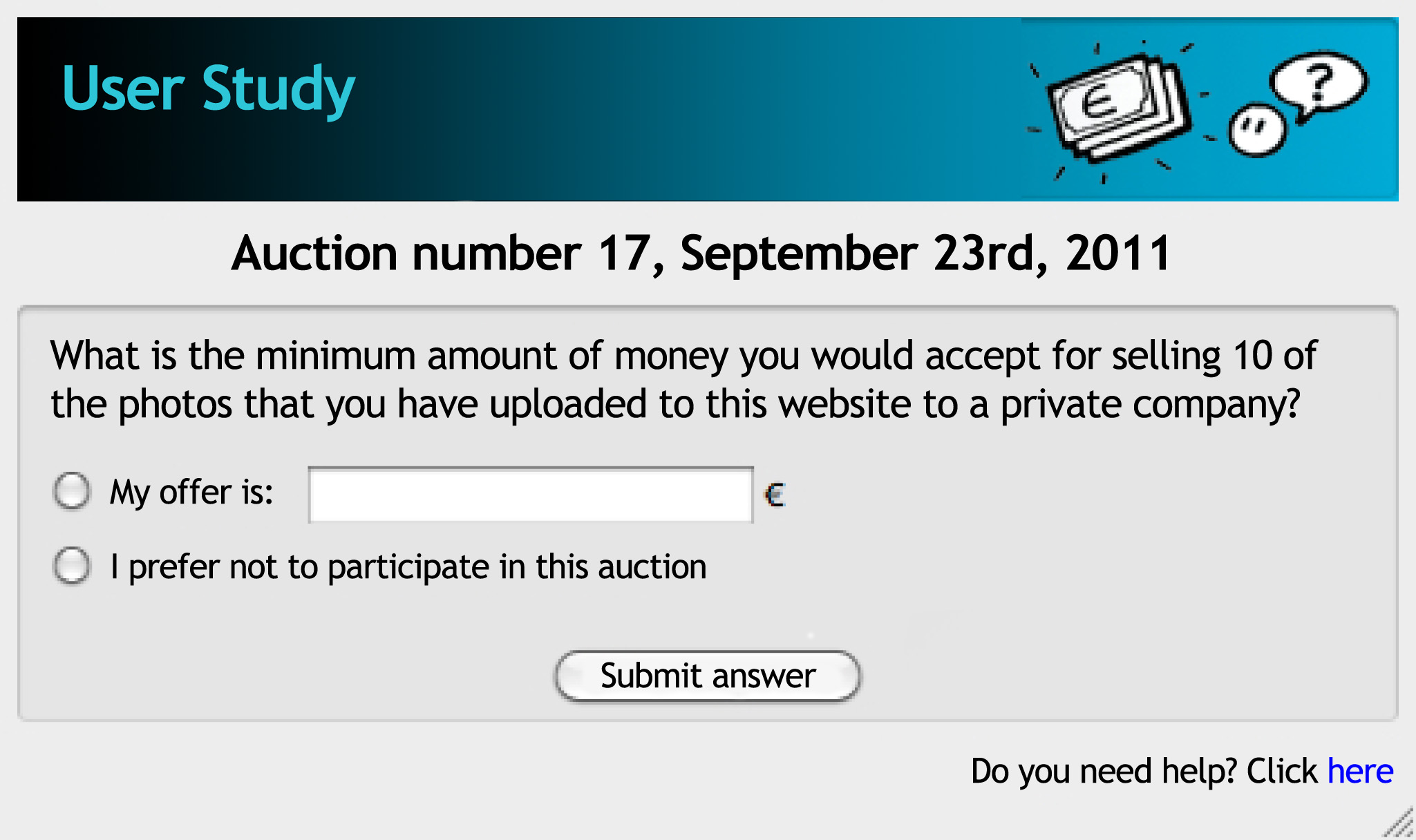}
    \caption{The auction popup. Each auction game was identified by a sequential
    number and a date. The participant had the option to either enter a bid or to
    not take part in the auction.}
    \label{fig:popup}
\end{figure}

We chose 8 categories (\textsc{Email}, \textsc{Entertainment},
\textsc{Finance}, \textsc{News}, \textsc{Search}, \textsc{Shopping},
\textsc{Social}, \textsc{Health}) to closely correspond to the 8 popular
categories that online ad-networks like
Doubleclick\footnote{Doubleclick has more than 8 major
categories, and more than 600 subcategories, but we chose 8 as a
good trade-off between obtaining detailed information without
annoying the user} use, as we are interested in the monetary aspect
of PI.

The plugin was able to sense when the user was changing context and
use this information to trigger specific questions to the users
about their perception of privacy and valuation of their private
information as explained in the following subsections.

\subsection{Participants}
Participants were recruited using a survey published via a major Web
portal in Spain.
From an initial pool of $279$ subjects, $168$ ($93$ male, $55\%$)
installed the Firefox\footnote{See
\url{http://mozilla.org/firefox}.} browser plugin and completed all
requirements of the study. All participants were users of the
Firefox browser and hence had it installed on their computer.
Participants' age ranged between $18$ and $58$ years old
($\bar{x}=31.83$, $s=8.15$). 
With respect to their educational level, $1\%$ had no level, $8\%$
finished primary school, $14\%$ did secondary school, $75\%$ had a
university graduate degree, and $2\%$ a post-graduate degree.
Socioeconomical status was also diverse: $28\%$ of the sample
informed their annual gross salary to be lower than \textgreek{\euro} 10K, $25\%$
said it was between \textgreek{\euro} 10K and 20K, for $22\%$ it was in the range
of \textgreek{\euro} 20K and 30K, $11\%$ between \textgreek{\euro} 30K and 40K, and $10\%$
reported earning more than \textgreek{\euro} 40K per year ($4\%$ preferred not
answering this question). All participants lived in
Spain
and the vast majority were of Spanish
nationality ($94\%$).

\subsection{Procedure}
The study ran for a period of $2$ months from mid-July to mid-September, 2011. Selected participants were
invited to take part in the study via email. The message contained a generic explanation of the
experiment where we mentioned we were interested in studying their privacy preferences when browsing and
a detailed explanation of install instructions of the browser plugin. We explained to participants that
the study consisted of three phases: (1) an initial week where the popups were inactive, (2) the actual study
that lasted $4$ weeks where popups were active, and then (3) the final questionnaire.

\begin{itemize}
\item During the initial week the plugin was silently recording the
browsing behavior of participants. We explained to the participants that we
were waiting for all the invitees to install the plugin before
starting the experiment. The information that was captured during
this phase was used to record the baseline browsing behavior to make
sure that our popups were not interfering with the way participants
normally browsed the internet.
In order to evaluate this, we extracted for every user the frequency
distribution across the visited sites -- we refer to this as the
user's \emph{fingerprint}. Each participant's fingerprint for the
first week was therefore compared against the second week's
fingerprint ($L2$ distance), when the pop-ups were activated.

\item During the experiment, the plugin displayed popups when the
participants were browsing the internet. The popups contained two
kind of questions: questions about their perceptions and
knowledge regarding monetization of PI (for RQ2) when browsing the particular website they were visiting,
and an auction (by way of a question) on the minimum value they would accept to sell a
particular piece of PI to us to use. We refer to the latter as the
auction game (described in detail in the next subsection). We were
deliberately vague about \emph{how} we were going to use their PI
for two reasons: (i) to realistically reflect the conditions that
exist today, where outside of large PI collectors like Google or Facebook,
there is little knowledge of how one's PI is
being used, (ii) not to bias the user by providing a specific use
case of their PI; for instance using PI for behavioral targeting can
be construed positively or negatively. However, in reality their information
was never used for any non-research purpose and it was discarded right after the study.
To avoid the popups being too
invasive the plugin was going to display at most one pop-up per
category per day. Also there was a \emph{minimum} delay of 10
minutes between any two pop-ups.

\item At the end of the experiment, we asked the participants to
fill in a post-study questionnaire in which we asked more detailed
questions on their knowledge of privacy threats, and who they
would trust with their PI. The analysis of these results is not
going to be part of this paper.
\end{itemize}

In terms of incentives, each user was given a gift card voucher
worth \textgreek{\euro} $10$ ($\sim$  $14$ USD). Also, we informed participants
that we were going to increase the value of their gift card with the
value of all the auctions they would have won during the time of the
experiment. Additionally, we specified in multiple occasions that
the maximum amount they could win during the experiment was \textgreek{\euro} $3000$
because we had a limited budget for the experiment.

Our ethical board and legal department approved the experiment. 
Participants were debriefed about what was being logged and instructed 
on how to disable temporarily or remove the plugin. Participants were 
free to leave the experiment at any time without consequences.

\begin{table*}[!t]
\vspace{-.2cm} \caption{Questions asked during the different phases of the study.}
\begin{tabular}{c}
\includegraphics[width=18cm]{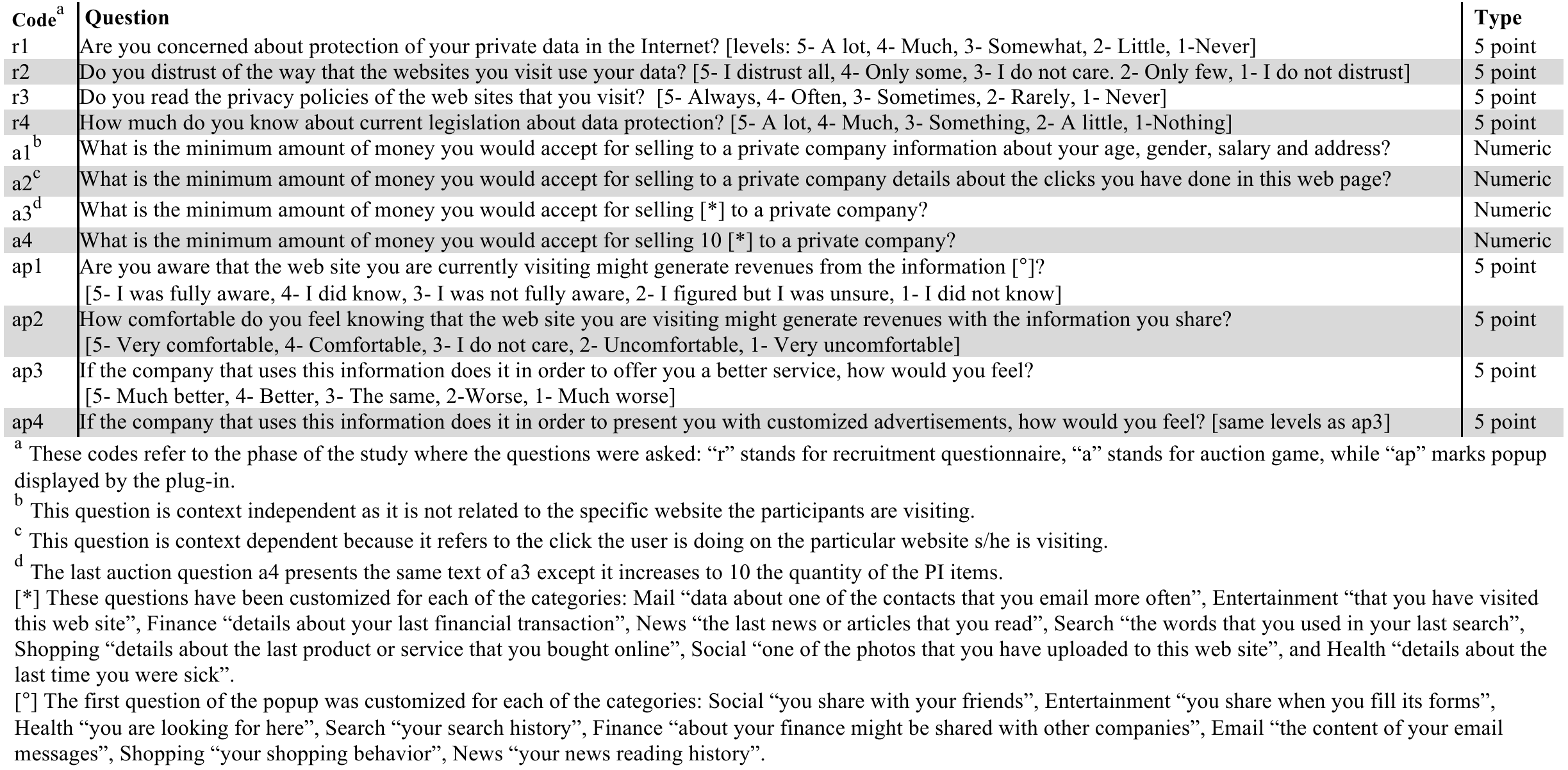}
\end{tabular}
\label{tab:questions}
\end{table*}

\subsection{Auction game}

In order to extract a concrete value that a user puts on her PI, we developed a simple game based on
the reverse second price auction. The reverse second price auction operates as follows: given a set of $k$ bids,
pick the lowest bidder as the winner, and pay that person the amount equivalent to the second lowest bid. This is
the opposite of what is used in online auctions like that of eBay. We chose this auction mechanism for the following
reasons: (i) this mechanism has the strong property of being truth telling; the best strategy for
participants in the auction is to be honest about their valuation~\cite{Hurwicz1978},
(ii) this mechanism has been used before for valuating location information~\cite{Ross-location}, (iii) this mechanism
is extremely simple and is a relatively easy mechanism to explain to users of our study.

We allowed positive amounts \emph{(including 0)} with as much as two decimals (for cents) as valid bids. We also gave the user a choice to not participate
in the auctions at all -- this was necessary to cover cases where users felt overwhelmed with participation
and more importantly, also the cases where users did not even want to disclose the fact that their PI is worth
a very high amount -- note that this by itself releases one bit of information.
In order to reinforce the notion that the user
will indeed part with their PI if they win, we had a second pop-up after the user enters an amount
that asks the user if they are sure that if they win that auction, they will part with the related PI.

For winners of the auction, we sent an email notifying them of their win, with following information:
their winning bid, and time of bid. We reinforced the message that as they won, we will
use their PI (exactly PI they bid on). Likewise, we sent a similar email to the losers, conveying that
as they lost, their PI will \emph{not} be used.
For all our communication with users, we used neutral language with regards to privacy, so as to not prime them
one way or another, following the findings in ~\cite{Braunstein2011}.

\subsection{Apparatus}
\input{design}

\subsection{Measures}

In terms of the measures that we used to answer our research
questions, Table~\ref{tab:questions} describes the most important
questions --coming from both the recruitment questionnaire and the
Experience Sampling-- that we presented the user during the study.

Questions r1-r4 are about gauging the knowledge of privacy related issues.
Questions related to the auctions were a1-a4, where a1 is a question
about PI related to off-line identity and is common across categories.
Questions a2-a4 are context dependent, with a2 about browsing information/history
and a3-a4 about category specific PI.

We chose to ask a2 as this is the information that most entities engaged in large scale
tracking across the web (like Google's DoubleClick or Facebook via their `Like' button) have access to,
and hence can monetize. These are often referred to as `third' parties.
Questions a3-a4 are category specific and in most cases, this PI is available only to the
service provider actually providing that service (photos on social networks, financial transactions,
purchase history on e-commerce sites etc.) These are referred to as publishers or `first' parties.

Questions ap1-ap4 were designed to understand if users are aware of
monetization of their PI by online entities. The first two questions (ap1,ap2)
had to do with knowledge and comfort levels of monetization, while ap3 has to do
with exchange of PI in return for enhanced services, for instance like recommendation
systems. Question ap4 is about personalized advertisements.

We called a1 \textit{context independent} because the PI we asked for does not relate 
to the website the user was visiting (although we presented the question multiple times 
using the rESM). The purpose of a1 was to assess the validity of our measures by 
contrasting with results from a2. Indeed, a2 and a3/a4 were \textit{context dependent}. 
But while the former asks about the same PI item across categories, the question in the 
latter is customized for each category of websites. Our goal was not to produce 
generalized estimates of context valuation but rather to understand whether online context 
had an influence on the valuation that people attach to certain types of PI.

\subsection{Statistical Analysis}

Nonparametric analysis was applied considering the ordinal nature of
some observed variables and that continuous variables did not follow
the normal distribution. Given that participants browsed web pages
in their natural environment without being enforced to visit sites
from all categories mapped in our study --thus promoting ecological
validity, our sample had several missing values across categories.
Removing subjects that did not provide information for all
categories --as they did not browse all types of web pages-- would
significantly reduce the generalization power of our results and
yield unrealistic findings based on the assumption that everybody
browses web pages from all categories considered in this study.
Therefore we opted for \emph{not} using related sample analysis.
Hence differences between median bid values (or Likert scale
measures) across categories were tested using the Kruskal-Wallis
test and the Mann-Whitney test whenever appropriate. Associations
between ordinal/interval variables were assessed using the
Spearman's Rho test. We considered widely accepted cutoff values
proposed by Cohen \cite{Cohen:1988} for determining the strength of
the correlations. The level of significance was taken as $p<.05$.

%% file: design.tex
In order to capture the browsing context of the auctions as well 
as the questions for understanding users' perception of PI exploitation, 
we developed a system consisting of two parts: a browser plugin and a web server 
that communicates with the plugin, sending configuration information 
to the plugin and receiving data from it.

\textbf{Firefox Plugin:} The plugin has three main tasks. First,
it captures and stores all browsing activity of the user.
This consists of the url, time of page access, and a unique ID we assigned to
the browser.  This data is stored on the local machine and sent to 
the server at regular intervals.  We do not capture events like file uploads, 
text highlighting etc. 

The second main task of the plugin was to categorize websites into 
one of the eight categories mentioned at the beginning of this section. In order to do this, we rely on a hard-coded list of 
$1184$ popular sites from different categories for Spain, gleaned from \texttt{alexa.com}. Although
some popular sites like Facebook can host content pertaining to health or entertainment, we 
hard-coded it to `Social' \footnote{Such a monolithic categorization does have limitations; large service providers
like Facebook or blogspot host content belonging to multiple categories. However, we consistenly pick the first category as put out by Alexa. This ensures
that we do not have any false positives -- Facebook will always be categorized as Social. We leave 
a detailed categorization mechanism to future work.}. For sites that are not on Alexa, we resolved them into categories 
by relying on a folkosonomy approach implemented in another browser plugin called Adnostic~\cite{adnostic}.
The details are provided in Toubiana et al~\cite{adnostic}, but the basic idea
is to perform a cosine similarity between the set of key-words present on the site the user visits
and a corpus of words that are associated with different categories. The category with the highest
similarity is used.

Third, the plugin has two independent pop-ups, as described earlier. The first plugin launched the auction mechanism 
and the other displayed questions related to privacy preferences. These are configured
to be switched on or off from the server. From a UI perspective, the pop-up displayed the
text of relevant auction question, with the type of PI in the auction in bold text, to highlight
what is actually being traded in the auction. There was a small box below the text where the user
could enter an amount, and there was a small radio button below the box where the user
can select to not participate in the auction (for reasons mentioned earlier).  

\textbf{Server:} We developed a simple, highly responsive webserver in Python that synced with 
the browser plugin at regular intervals. The server accepts data (bids, responses to the questions) 
from the plugin and stores it in a sqlite database.  The main function of the server
is to run auctions. For each category and for each type (there are 4 types per category),
we set an auction to run once 20 bids are in. We pooled all these auctions and ran them once daily,
in the morning. This was all automated.  We sent out results to participants (winners and losers) 
via emails.

%% file: results.tex

We summarize the main results obtained with the user study
towards addressing our two research questions.

\subsection{Effect of pop-ups on browsing behavior}

We found little deviation between participants' first week's
fingerprints -- baseline -- and their fingerprints for the second
week of the study -- when pop-ups were turned on. Specifically, only
three users ($2\%$ of the sample) presented higher browsing behavior
deviation and reported being on vacation during the second week,
thus explaining why they used their browser sparsely. These findings
indicate that users \emph{did not} deviate from their `normal'
browsing behavior when participating in the study.

\subsection{Results for RQ1}


Findings presented herein shed light on the value that users of web
services attribute to the information they share online. First we
briefly summarize results for the winning bids ($n=40$), followed by
more generic results comprising the whole sample ($N=168$).

\emph{Winning bids and pay-outs.} Considering the $40$ subjects that
won at least one auction, their median winning bid was of $5$ cents
of Euro ($min=0, \bar{x}=0.19, max=2.29$). Even though we allowed a
bid of $0$ as a valid bid, only seven winners bid $0$ on $11$
occasions, out of 5000+ bids.

The other winners' bids were strictly positive. Finally, as we used
the reverse second price auction, the median payout was actually 45
cents of Euro ($min=0.01, \bar{x}=0.65, max=5.69$).

\emph{Representativeness of categories.} Next we look into the
bidding behavior of the whole sample ($N=168$) while browsing
websites as they map to each of the $8$ categories and also in relation
to the nature of the information being sold (see questions
\emph{a1-a4} in Table~\ref{tab:questions}). Overall, participants
visited websites from all of the eight categories,
\textsc{Health} being the least visited category (\textsc{Search}=$82\%$,
\textsc{Entertainment}=$82\%$, \textsc{Social}=$78\%$,
\textsc{News}=$76\%$, \textsc{Finance}=$75\%$,
\textsc{Shopping}=$75\%$, \textsc{Email}=$64\%$,
\textsc{Health}=$2\%$). Given the lack of representativeness for the
number of subjects visiting health related web pages, we therefore
decided to consider only seven categories when comparing
participants' bids and other relevant measures across categories.

\emph{Bids on context independent PI.} With respect to selling their
PI that is related to their offline identity (\emph{i.e.}, age,
gender, address and bank balance; see question \emph{a1} in
Table~\ref{tab:questions}), we found no significant difference among
participants' median bid values across categories ($p=.702$). Note
that this result was somewhat expected as question \emph{a1} was
context independent -- no mention was made to selling the
participants' PI to an entity related to the website they were
browsing. The overall median bid value across categories was \textgreek{\euro}
 25.

\emph{Bids on context dependent PI.} When probed about selling
clicks they performed on a given web page (see question \emph{a2} in
Table~\ref{tab:questions}), which represents their browsing behavior, participants' median bids were not
significantly different across categories ($p=.569$). In this case,
the overall median bid value was \textgreek{\euro}
 7\footnote{This is approximately the value of a BigMac meal in Spain, 
circa 2011. Hence the title of this paper.}.

Median bid values for highly category specific PI -- as captured by
questions \emph{a3} and \emph{a4} in Table~\ref{tab:questions} --
revealed significant differences across categories ($p<.001$). The
highest median bid values were from categories \textsc{Finance}
($\tilde{x}=15.5$), \textsc{Social} ($\tilde{x}=12$), and
\textsc{Email} ($\tilde{x}=6$), being \textsc{Finance} similar to
the latter two categories ($p=.31$ and $p=.09$ respectively) and
significantly different to the remaining categories
(\textsc{Shopping}=$5$, \textsc{News}=$2$,
\textsc{Entertainment}=$2$, \textsc{Search}=$2$; $p<.001$).

\emph{Bulk PI effect.} We verified no significant difference between
the median bid value for all categories in question \emph{a3}
($\tilde{x}_{a3}=5$) and in question \emph{a4} ($\tilde{x}_{a4}=5$,
$p=.59$). This finding indicates that the amount of information
being sold was not a factor for participants when placing their
bids, as they valued one piece of information (question \emph{a3})
and 10 pieces of information (question \emph{a4}) in a similar way.

Table~\ref{tab:bids} summarizes the most relevant descriptive
statistics of median bid values per category.

\begin{table*}[t!]
\vspace{-.2cm} \caption{Median bid values per category calculated
from participants' median bids in each category (1st and 3rd
quartiles shown between brackets). Similarity across categories
indicated by $p$-values. See Table~\ref{tab:questions} for details
on questions \emph{a1-a4}.}
\begin{tabular}{c}
\includegraphics[width=17.5cm]{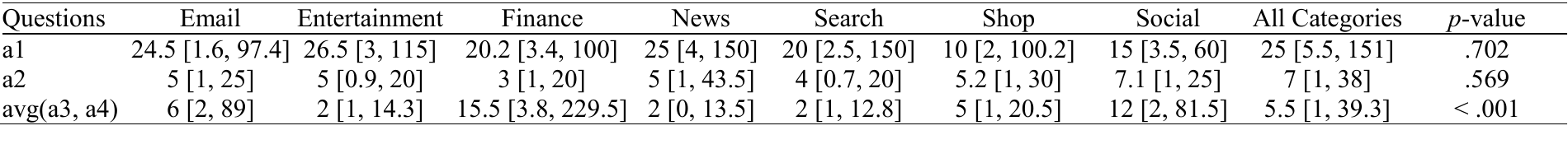}
\end{tabular}
\label{tab:bids}
\end{table*}

\emph{Relationship between bids, demographics, and privacy.} We
further looked into significant associations between variables
captured in the recruitment questionnaire and the participants'
bids. Our findings reveal a medium negative correlation between
participants' age and their median bid values for question
Social-\emph{a3} ($n=64$, $\rho={-.276}$, $p=.03$). Similarly, age
is negatively correlated to the combination of questions
Social-\emph{a3} and Social-\emph{a4} ($n=69$, $\rho={-.287}$,
$p=.02$), thus providing evidence that the older people are, the
lower they tend to bid on photos they share online. Furthermore, we
found a medium positive association between gender and median bids
for question Email-\emph{a3} ($n=45$, $\rho={.333}$, $p=.03$). This
result indicates that men might bid higher than women on information
related to their email contacts. Correlations between income levels
and bid values were not significant. Finally, we found medium
negative correlations between participants' education level and
their median bid values for question \emph{a2} in most categories
(\textsc{Entertainment}: $\rho=-.277$, \textsc{Finance}:
$\rho=-.282$, \textsc{Search}: $\rho=-.235$, \textsc{Shopping}:
$\rho=-.32$).

We also correlated bid values with responses provided to
privacy-relevant questions in the recruitment questionnaire. 
Positive medium correlations were found between being worried about
online data protection and higher bids on context independent PI
(question \emph{a1}, \textsc{Entertainment}: $\rho=.252$,
\textsc{Finance}: $\rho=.278$, \textsc{Search}: $\rho=.23$).

\subsection{Results for RQ2}

Results presented in this subsection contribute to the understanding
of how users' perceive the economic usage of their PI by online
service providers. Note that we considered only the first answers
that participants gave to questions \emph{ap1-ap4} per category.
This decision guaranteed that their initial opinion would be taken
into account instead of a -- potentially -- biased opinion due to
the effect of long exposure to the study.

\emph{Knowledge of PI-based monetization.} Participants were aware
that PI shared in a particular web site could be used to generate
revenue (question \emph{ap1}, $\tilde{x}=4$, $q1=2$, $q3=4$).
Moreover, no significant difference was found between median ratings
across categories ($p=.107$). This finding suggests that knowledge
of PI-based monetization is related to Internet services in general
and not to a particular set of services.

\emph{Comfort with PI-based monetization.} In question \emph{ap2},
participants revealed how comfortable they were with web sites
extracting revenue out of their PI. With a median rating of 2
($q1=2$, $q3=3$), they reported being uncomfortable with it, and
this feeling was shared across categories as no significant
difference between participants' median ratings per category could
be found ($p=.429$). From this finding, we conclude that the act of
monetizing from users' PI is what generally makes people
uncomfortable, and not the type of online service providers the
revenue will go to (\emph{e.g.}, finance, search, \emph{etc.}).

\emph{Improving services with PI.} Although not comfortable with
their PI being monetized, participants pointed out that they would
like online companies to improve their web services using their PI
(question \emph{ap3}, $\tilde{x}=4$, $q1=3$, $q3=4$). No significant
difference was found between participants' median ratings across
categories ($p=.869$).

\emph{PI-based publicity/ads.} Finally, subjects were indifferent with
regards to online service providers making personalized publicity/ads by
using their PI (question \emph{ap4}, $\tilde{x}=3$, $q1=3$, $q3=4$).
Once again no significant difference could be identified between
participants' median ratings across categories ($p=.686$). This
finding suggests that personalized ads from web services belonging
to different categories generally have neither a negative nor a
positive impact on people.

%% file: discuss.tex
\subsection{Users value offline PI more and online PI less}
If we consider the results for a1 (Sec:Results)
users consistently bid high values for their offline PI like age, gender,
address and financial status; pieces of PI that form their
off-line identity, to trade with online entities.
Likewise, users attach lower value (relatively) to
a2, a3 and a4, PI that
mostly has to do with their online behavior (a2 is exclusively about
browsing history, the other two are about online transactions).
Digging deeper, we also note that users tend to value category-specific PI (a3 and a4)
on \textsc{Finance} and \textsc{Social}, categories that are more explicitly
intertwined with one's off-line identity, more than \textsc{Search} and \textsc{News}.

This may seem contradictory to the conjecture put forth in ~\cite{Acquisti2004-sec},
where the author claims users act ``myopically when it comes to their off-line identity
even when they might be acting strategically for what relates to their on-line identity."
The author puts forward the need for immediate gratification and hyperbolic discounting
of future risks of revealing PI pertaining to off-line identity as possible explanations.

First, we do not believe our result is contradictory -- note that we are comparing \emph{economic}
value attached to off-line PI as opposed to PI created online, not disclosure strategies.
Second, we conjecture that the difference in valuation exists
because of lack of awareness. Off-line PI is easier to valuate as it is more explicit.
It is harder to understand the implications of being continuously tracked and then the collected
PI (browsing information) being data mined to produce
a unique profile and be linked to an off-line identity~\cite{Goel, url:rapleaf}.
As a consequence, users value such PI less.

\subsection{Users do not distinguish between quantity of PI, but type}

We compared the median bid values for a3 and a4 across categories and found little or no
difference. These two auction questions differ only in quantity of information being traded, with
the type of PI and the context remaining the same. As reported above, there are significant
differences between type (\textsc{Finance} and \textsc{Social} being higher than \textsc{Search}, \textsc{Shopping} etc.)

We correlated the values with demographic information as well as the responses to the
privacy related questions (r1-r4). We found little correlation.
A possible conjecture can be on the lines of what is reported in ~\cite{cvrcek2006}, that users
factor in diminishing returns of more information in their valuation -- although
we have no evidence to support or refute this conjecture.

\subsection{Older users less concerned about online PI}

When we correlated bid values against demographics, a high (negative) correlation
occurred between age and category specific PI on \textsc{Social, Entertainment} and \textsc{News},
and more so while valuating bulk information (a4). For \textsc{Social},
this can be linked to the fact most older users
do not use online social networks, let alone upload photos
to online social networks\footnote{http://www.comscoredatamine.com/2010/09/visitor-demographics-to-facebook-com/}.

This result is in contrast to previous work that stated that older users are generally more concerned
about their privacy, while being online~\cite{Paine}. We believe our results underscore the point
made by Acquisti et al~\cite{Acquisti2004-sec}, that there are often differences between stated privacy preferences and
actual behavior.

\subsection{Users do not like monetization of their PI}
\label{subsec:nolike}

When we consider the results of our analysis on the responses to
the questions we posed to users, the following
trends stand out.  First of all, users are overwhelmingly negative when
it comes to their PI being used for monetization by entities (ap2), despite knowing
that online entities collect and use their PI for monetization (ap1). In addition, they
prefer their PI to be used for improving the services they are offered (ap3), across all categories.
On the one hand, these results are expected -- the former deals
with monetization of a good (PI) that users probably perceive as theirs, while
the users view the latter as a positive outcome of their PI being exploited.

The combined results possibly point to the fact that users are unaware of the
functioning of the ecosystem in place -- they do not perceive that the services they get
for `free' (storage in Gmail, Google search, Facebook etc.)
actually are expensive (large datacenters, equipment and bandwidth costs) and
while users are aware of their PI being monetized, they are possibly not
aware that large parts of that monetization goes towards providing them with
a `free' service.  It was reported in~\cite{acquisti-beh}, where the authors
claim that users are more sensitive about their privacy and PI when they feel
that service providers are unfairly gaining from the use of PI. This unfairness feeling
can be due to lack of awareness.

Second, users are indifferent when it comes to the use of the PI to
send them personalized ads (ap4), again across categories. This is somewhat in contrast to
results in~\cite{advert} where the authors report that 64\% of the survey respondents (all Americans)
find behavioral targeting invasive. The differences between our results
and theirs can be due to cultural differences (our sample consists mainly of people from Spain) 
and/or methodological differences -- we used experience sampling to capture the context,
while the results reported in~\cite{advert} were gathered via surveys.

%% file: implications.tex

Our study has direct implications on the monetization of personal information (PI) online. As the focus of the study has been towards understanding the \emph{economic} aspects of PI, we believe the findings can help in the following future research topics and new offerings. We propose three major implications.

\subsection{Markets for PI}
Recent years has seen rise of interest in online privacy concerns and collection and exploitation of PI from multiple quarters -- mainstream press (WSJ's `What they know' series~\cite{url:what} etc.), research on how PI is used (behavioral targeting~\cite{advert}, price discrimination~\cite{erosion} etc.) and move towards regulatory actions~\cite{url:privacybill}. Irrespective of the specific message, what all sources agree on is that data collection (online and mobile) is \emph{increasing} and this increase is related to the rise of the `free' model of providing online services. Hence, on one side you have entities like Google who have stated that they want to move up to the `creepy' line~\cite{url:creepy} on accessing and using PI, while users are resorting to measures like Do-not-track\footnote{http://donottrack.us/} etc. to prevent data collection, leading to an impasse.

As mentioned in the Introduction, the current economic system around online PI is a two-sided market, 
with sellers/providers of PI on one end, and buyers/exploiters of PI on the other, 
with the network (Internet) in the middle. Looking at the problem this way, one solution to the impasse 
above is to have a market for personal information, where users can decide to sell the PI of their choice, legally, 
to online service providers, who will in turn exploit the PI they have purchased. As users have a choice in 
deciding what PI about them gets traded, and receive monetary compensation, this will decrease privacy concerns. 
The general attitude of participants in taking part in the auctions and their willingness to sell their PI point to this implication.
In addition, the fact that users are aware of their PI being monetized but are not happy with fact, can point to 
the notion that users feel they are not being adequately compensated in today's ecosystem and an open market can help
in addressing this issue.
This idea has roots in Laudon~\cite{laudon} and has recently gained traction for online PI~\cite{GhoshR11, chris-hotnets}.

The results in this paper provide the first empirical foundation for such a market by demonstrating how users 
value different types of PI in terms of different types of interactions they perform while online, as well as in context. 
The prices can be taken to be the reserve prices\footnote{http://en.wikipedia.org/wiki/Reservation\_price} that users will be 
willing to accept to part with their PI. Likewise, we have seen that different types of interactions and PI 
have different valuations (photos in social networks \textit{vs.}~online purchase history). These differences can 
be used by service providers to strategically target different types of PI. That is, service providers can decide that 
it may not be economically viable to purchase offline PI about users, while using PI about \textsc{Search} or \textsc{Entertainment} 
might be more economically sound. This, in turn can also lead to a decrease in privacy violations. 
From a research perspective, the findings in our paper can be used as inputs to drive models to better understand the
ecosystem.

A simple market can be built around selling one's personal photos. Consider the scenario where the 
user has uploaded photos to a site. The user can select which photos can be `sold'; used for
some commercial purpose by the site. The site compensates the user after adjusting for hosting costs. 
Moreover, the user can sell the same set of photos to multiple sites, as she sees fits. 

\subsection{Transparency on monetization of PI}
From one of the findings reported in the Discussion section, while users have knowledge of their PI being collected, they 
are not comfortable about their PI being monetized. This lack of awareness also plays out in valuations --while 
offline PI and certain types of online PI like photos, financial transactions have high valuations, presence of the user 
on different sites are valued very low. This is interesting as a behavioral profile can be constructed just by tracking 
users across sites (via cookies etc) and this profile can be used to identify users and be monetized~\cite{beh}. We believe that 
most privacy concerns that arise is due to lack of awareness of precisely this fact --that PI is being 
monetized (participants knew their PI could be monetized by entertainment and search related websites, 
but not for the other categories). 

The findings reported in this paper indicate that if online service providers are explicit and up front about the fact 
that they provide a service (email, video streaming, a social network, etc) for free and in return collect and monetize PI, 
along with details on the specific types of PI they collect, the privacy concerns of most users will be tempered. 
Long privacy policies written in complicated legalese that are seldom effective~\cite{Jensen2004}, can be dispensed with. 
For example, we can think about agreements that could expose the amount of money required to run the service the user is signing 
up for and how the revenues generated by exploiting PI help cover those costs. Additionally, we can think about alternative 
business models where the user has the option to pay for the service that s/he is signing up for either with his/her PI or 
with real money.

\subsection{Bulk data mechanism}
A final implication for design is related to the indifference in valuation 
for bulk quantity of data. Specifically, participants assigned a similar value to a certain piece of PI as to
$10$ pieces of the same information. This has a direct consequence for the design of 
selling PI in the markets (described above). In fact, it does not make sense to implement mechanisms for the sale of a 
single piece of information. Rather, it makes more sense --according to these results-- to design solutions that would 
allow interested users to sell a bulk amount of PI. For instance, such a mechanism could be presented during 
registration to a new service and extended for bulk amounts of PI that the user will be sharing throughout 
the use of the service. The effect of such a design could be two fold:  on one hand it would minimize 
the user's effort and mental load, while on the other hand it would maximize the effectiveness 
of the service provider's budget expenditure.

%% file: conc.tex
Our study focused on two questions. The first has to do with understanding 
the monetary value that users put on different types of PI in an online context. The second 
has to do with understanding general attitudes towards collection and exploitation of personal information, again in context. 

Previous literature has shown that privacy valuation is a difficult problem, as it is affected by a number of 
technical, legal, social and psychological factors, amongst others, that lead to inconsistencies between 
what people say and what they actually do. We consider that our approach, employing a refined Experience Sampling Method,
paired with a truth-telling auction mechanism allowed us to overcome the existing gap between reported preferences 
and actual behavior regarding online privacy.

We found that users give more importance to PI related to their offline identities than to PI that is related to
their online behavior. They mostly do not care about the 
amount of PI released but they do care about its type. Finally, even though people 
consider that the use of their personal information for improving service, 
they do not like their information to be used to generate revenues.

The need to be connected to the Internet seems to be constantly pushing privacy boundaries, and 
we should try to understand what it means both for users who are putting more of their lives online, 
and for entities interested in monetizing that fact. Though it is difficult to address all these factors 
in one single study, we believe our work will help in understanding the underlying mechanics at work,
from an economic perspective.